\documentclass[12pt]{article}
\pdfoutput=1

\usepackage{color}
\usepackage{amsmath}
\usepackage{amsfonts}
\usepackage{amssymb}
\usepackage{graphicx}
\usepackage{subfig}             

\usepackage[english]{babel}
\usepackage[applemac]{inputenc}
\usepackage{braket}

\usepackage{cite}

\usepackage[	colorlinks=true,	citecolor=black,	linkcolor=black,	urlcolor=blue,hypertexnames=false]{hyperref}

\def\beq{\begin{eqnarray}}
\def\eeq{\end{eqnarray}}
\def\bean{\begin{equation*}}
\def\eean{\end{equation*}}
\newcommand{\arXiv}[2]{\href{http://arxiv.org/pdf/#1}{{\tt #2/#1}}}
\newcommand{\arXivold}[1]{\href{http://arxiv.org/pdf/#1}{{\tt #1}}}

\numberwithin{equation}{section} 

\newcommand{\RN}[1]{%
  \textup{\uppercase\expandafter{\romannumeral#1}}%
}

\newcommand{\bsp}{\begin{split}}
\newcommand{\esp}{\end{split}}

\newcommand{\bag}{\begin{align}}
\newcommand{\eag}{\end{align}}

\definecolor{MyBlue}{rgb}{0.1,0.1,0.8}

\begin{document}
\begin{titlepage}

\begin{center}

	{	
		\LARGE \bf 
		Composite Twin Dark Matter
	}
	
\end{center}
	\vskip .3cm
	
	\renewcommand*{\thefootnote}{\fnsymbol{footnote}}

%

\begin{center} 
{\bf \  John Terning\footnote{\tt \scriptsize
		  \href{mailto:jterning@gmail.com}{jterning@gmail.com}} 
		  },
{\bf \  Christopher B. Verhaaren\footnote{\tt \scriptsize
		  \href{mailto:cbverhaaren@ucdavis.edu}{cbverhaaren@ucdavis.edu}}
}, and 
{\bf \  Kyle Zora\footnote{\tt \scriptsize
		 \href{mailto:kezora@ucdavis.edu}{kezora@ucdavis.edu}	}	 
}

\end{center}

	\renewcommand{\thefootnote}{\arabic{footnote}}
	\setcounter{footnote}{0}


\begin{center} 

	{\it Department of Physics, University of California Davis\\One Shields Ave., Davis, CA 95616}

\end{center}


\centerline{\large\bf Abstract}

\begin{quote}
We consider Fraternal Twin Higgs models where the twin bottom quark, $b^\prime$, is much heavier than the
twin confinement scale. In this limit aspects of quark bound states, like the mass and binding energy, can be accurately calculated. We show that in this regime, dark matter can be primarily made of twin baryons containing $b^\prime b^\prime b^\prime$ or, when twin hypercharge is gauged, twin atoms, composed of a baryon bound to a twin $\tau'$ lepton. We find that there are significant regions of parameter space which are allowed by current constraints but within the realm of detection in the near future. The case with twin atoms can alleviate the tension between dark matter properties inferred from dwarf galaxies and clusters.
\end{quote}

\end{titlepage}

\section{Introduction\label{s.intro}}
One of the most compelling motivations to search for new particles and interactions beyond the standard model (SM) is the so-called dark matter (DM) that makes up 80\% of the matter density of the universe. The indirect evidence for DM is overwhelming \cite{Rubin:1978kmz,Clowe:2006eq} but neither the mechanism for producing the inferred density of DM particles nor its mass or non-gravitational interactions have been experimentally identified. The paradigm of asymmetric dark matter (ADM) \cite{Kaplan:1991ah} is motivated by the observation $\Omega_\text{DM}\simeq 5\Omega_\text{B}$, where $\Omega_\text{DM}$ and $\Omega_\text{B}$ are the DM and baryonic mass densities respectively. If matter/antimatter asymmetries in both the visible and dark sectors have a common origin, the similarity in their mass densities is natural, rather than a miraculous conspiracy between two a priori independent processes. In particular, we have
\beq
\frac{\Omega_\text{DM}}{\Omega_\text{B}}=\frac{\eta_\text{DM}}{\eta_\text{B}}\frac{m_\text{DM}}{m_N}\,,\label{e.adm}
\eeq
where $\eta_\text{DM}$ ($\eta_\text{B}$) sets the dark matter (baryon) asymmetry and $m_N$ is the nucleon mass. Clearly, ADM is even more appealing in models where there is some symmetry between the dark and visible sectors, as in Mirror world scenarios~\cite{Blinnikov:1982eh,Carlson:1987si,Foot:1991bp}, which can ensure $\eta_\text{DM}\sim\eta_\text{B}$ and $m_\text{DM}\sim m_N$. 

Explaining the hierarchy between the weak scale and the higher scales associated with modifications of the SM, including the Planck scale, has long guided explorations beyond the SM. Within the variety of possibilities that have been considered, the paradigm of neutral naturalness encapsulates those frameworks which explain the little hierarchy, between the weak scale and a few TeV, through a new symmetry, but whose partner quarks do not carry SM color~\cite{Chacko:2005pe,Barbieri:2005ri,Chacko:2005vw,Burdman:2006tz,Cai:2008au,Poland:2008ev,Craig:2014aea,Batell:2015aha,Serra:2017poj,Csaki:2017jby,Cohen:2018mgv,Cheng:2018gvu}. The first and most studied realization of this idea is the twin Higgs~\cite{Chacko:2005pe} scenario. It remains a future target of collider tests~\cite{Burdman:2014zta,Buttazzo:2015bka,Curtin:2015fna,Curtin:2015bka,Cheng:2015buv,Cheng:2016uqk,Contino:2017moj,Ahmed:2017psb,Chacko:2017xpd,Buttazzo:2018qqp,Bishara:2018sgl,Kilic:2018sew,Alipour-fard:2018mre} and may also have connections to neutrinos~\cite{Bai:2015ztj} and flavor~\cite{Csaki:2015gfd,Barbieri:2017opf}. In addition, the twin Higgs is a simple, concrete framework for thinking about dark matter sectors with a rich variety of particles and interactions. 

In twin Higgs constructions, the SM particle content is doubled, making a visible sector and a twin sector, which are related to each other by a discrete $Z_2$ symmetry. This means that the twin sector has the same gauge structure of the SM, but the gauge groups are distinct, so that the fields of one sector are gauge singlets of the other. In addition, the scalar potential is approximately invariant under an $SU(4)$ global symmetry~\cite{Chacko:2005pe}. When this symmetry is spontaneously broken by a vacuum expectation value (VEV) $f$ down to $SU(3)$, seven pseudo-Nambu-Goldstone bosons result. Six are eaten by the $SU(2)$ gauge symmetries in either sector, leaving one physical Higgs boson. To satisfy experimental bounds on the couplings of the Higgs to SM fields, the discrete symmetry must be softly broken, such that the VEV in the SM sector $v=246$ GeV is a few times smaller than $f$, $f/v\gtrsim 3$~\cite{Burdman:2014zta}. The larger value of $f$ leads to masses in the twin sector being raised beyond their SM counterparts
\beq
m_\text{Twin}=\frac{f}{v}m_\text{SM}~.
\eeq
If the ratio $f/v$ is taken to be too large then the twin top-quark mass $m_{t'}$ becomes much heavier than the SM top quark, signaling a fine-tuning, for instance if $m_{t'}=1$ TeV the theory is tuned to about 10\%~\cite{Burdman:2014zta}. 

This mirror twin Higgs construction includes many new light states in the hidden sector. This seems at odds with cosmological measurements of the number of light relativistic species, in particular the CMB measurement of $\Delta N_\text{eff}\lesssim$0.3 at 95\% confidence~\cite{Aghanim:2018eyx}. It has been shown that such cosmological tensions can be overcome~\cite{Barbieri:2016zxn,Chacko:2016hvu,Craig:2016lyx,Csaki:2017spo} while continued analysis of cosmological data may reveal signatures of a twin-like structure~\cite{Freytsis:2016dgf,Prilepina:2016rlq,Chacko:2018vss,Fujikura:2018duw,Li:2019ulz}. Models of baryogenesis~\cite{Farina:2016ndq} as well as dark matter~\cite{Garcia:2015loa,Craig:2015xla,Garcia:2015toa,Farina:2015uea,Hochberg:2018vdo,Cheng:2018vaj} have also been explored.

A simple way to relieve tension with $\Delta N_\text{eff}$ is to remove the light degrees of freedom from the twin sector. In the Fraternal Twin Higgs model~\cite{Craig:2015pha} only the third generation of quarks and leptons are twinned.\footnote{The vector-like twin Higgs~\cite{Craig:2016kue} model provides vector-like masses to the third generation quarks, removing the need for twin leptons to cancel gauge anomalies. } This construction strives for minimal constraints from naturalness on the twin sector. Therefore, only the twin top quark $t'$ needs to be a nearly exact twin, that is with the same Yukawa coupling to the Higgs. Other twin fields like the twin bottom quark, $b'$, and twin tau, $\tau'$, can differ significantly from the mirror model expectation, as long as their Yukawas do not become close to top Yukawa size. Similarly, the twin strong coupling $\alpha_s'$ may not be exactly equal to the SM value at the cutoff of a few TeV. If it is too different, more than a few tens of percent, then the two-loop running of the top Yukawa diverges from the SM value, spoiling the cancellation of quadratic divergences. Of course, the different particle content at low energies means the twin QCD scale $\Lambda_{\text{QCD}^\prime}$ can be considerably larger than the SM confining scale, often taking values of a few GeV. However, if the twin coupling is smaller than the SM value at the cutoff, confinement scales of a few hundred MeV can result while preserving naturalness.

This fraternal construction has been previously explored as an interesting candidate for ADM \cite{Garcia:2015toa}, but only in the regime where the $b'$ mass is comparable or smaller than $\Lambda_{\text{QCD}^\prime}$.\footnote{In~\cite{Farina:2015uea} the ADM construction is studied in the mirror twin Higgs model.} In the $m_{b'}\ll \Lambda_{\text{QCD}'}$ limit it is easy to estimate the mass of the stable baryon composed of twin bottom quarks leading to a robust analysis of the proposed dark sector, with or without gauging twin hypercharge. Here we explore the opposite regime where of $m_{b'}\gg\Lambda_{\text{QCD}^\prime}$. In this case the physics of the composite baryon is simpler; it can be understood using non-relativistic quantum mechanics. However, it requires slightly more effort than the previous analysis to obtain an accurate mass estimate. In the next section we review the calculation of the baryon mass, and then turn in Sec.~\ref{s.Bconst} to the constraints on these baryons as an ADM candidate. In Sec.~\ref{s.DarkAtoms} we consider the case that twin hypercharge is gauged so that neutral twin atoms may form out of the twin baryon and the twin $\tau'$, which provides another interesting DM candidate. There we also comment on a possible resolution of the tension between dark matter self-interaction cross sections inferred from dwarf galaxies and the bullet cluster
\cite{Chu:2018faw,Kaplinghat:2015aga}.


\section{Baryon Masses\label{s.barymass}}
In the heavy $b^\prime$ regime, $m_{b^\prime}\gg \Lambda_{\text{QCD}^\prime}$, we can treat the QCD$^\prime$ coupling, $\alpha_s^\prime=g_3'^2/(4\pi)$, as perturbative at scales of the size of the bound states. The lightest color neutral baryon, the analog of the spin 3/2 $\Delta$ baryon, is then simply a Coulombic bound state of three identical heavy quarks, as long as the Bohr radius $\sim 1/(\alpha_s' m_{b'})$ is much smaller than the confinement length $\sim 1/\Lambda_{\text{QCD}^\prime}$. We can then calculate approximate $\Delta^\prime$ masses using a non-relativistic Hamiltonian:
\begin{equation}
H=\sum_{i=1}^3 \frac{\mathbf{p}^2_i}{2m_{b^\prime}}-\frac{\mathbf{P}_\text{CM}^2}{2M}+\left(\frac{1}{9}\alpha^\prime-\frac{2}{3}\alpha^\prime_s\right)\sum_{i>j}^3\frac{1}{r_{ij}},
\end{equation}
where $\mathbf{P}_\text{CM}$ is the momentum of the center of mass, $M$ is the total mass, $\alpha^\prime$ is the coupling for twin QED, and $r_{ij} = |\mathbf{r}_i-\mathbf{r}_j|$ is the distance between the $i$th and $j$th quarks. The factor of $\frac{2}{3}$ arises from the Casimir of twin color generators $T^a T^a$ evaluated in the color singlet state. The baryon ground state can be roughly approximated by the wavefunction
\begin{equation}
\Psi(\mathbf{r}_1,\mathbf{r}_2,\mathbf{r}_3)=8a^{9/2}\exp\left[-a(r_1+r_2+r_3)\right],
\end{equation}
where $a$ is a variational parameter. Using values $\alpha^\prime=1/137$ and $\alpha_s^\prime=0.15$, the variational binding energy is $E\approx -0.00865m_{b^\prime}$, so the baryon mass is about three times the mass of its heavy constituent quarks, as expected.

A more precise solution is obtained by using the stochastic variational method~\cite{Varga:1995dm}. We use a correlated Gaussian (CG) basis
\begin{equation}
\phi(\mathbf{r}_1,\mathbf{r}_2,\mathbf{r}_3)=\exp\left(-\frac{1}{2}\sum_{i,j=1}^2 A_{ij}\mathbf{x}_i\cdot\mathbf{x_j}\right),
\end{equation}
where $A$ is a positive definite matrix and the $\mathbf{x}_i$ are the Jacobi coordinates of the 3-quark system. In particular,
\begin{equation}
\mathbf{x} = \mathbf{Ur},
\end{equation}
with
\begin{equation}
\mathbf{U} = \begin{pmatrix}
1 & -1 & 0 \\
1/2 & 1/2 & -1 \\
1/3 & 1/3 & 1/3
\end{pmatrix}.
\end{equation}
Using Jacobi coordinates eliminates our need for a center of mass coordinate, which speeds up the calculation. 

Since the three quark color singlet is antisymmetric, and the complete 3-quark wavefunction must be antisymmetric overall, we symmetrize the spatial wavefunction with
\begin{equation}
S = \frac{1}{\sqrt{3!}}\sum_{n}\mathbf{P_n},
\end{equation} 
where the $\mathbf{P_n}$ are permutation elements of the symmetric group $\mathbf{S}_3$. This operator commutes with the Hamiltonian and has the property
\begin{equation}
S^\dagger S = \sqrt{3!}S\,.
\end{equation}

We find a basis set, by first generating a set of random, positive definite matrices $A$ and select the one which minimizes the binding energy. Additional matrices are added to the basis set when doing so reduces the binding energy by at least a specified amount. 
The energy is then calculated using the  basis set of wavefunctions $|\phi_i\rangle$ to produce a variational wavefunction 
\begin{equation}
|\Phi\rangle=\sum_i c_i|\phi_i\rangle.
\end{equation}
We then solve the generalized eigenvalue problem for the vector $c_i$
\begin{equation}
\mathbf{\mathcal{H}c}=E\mathbf{\mathcal{N}c},
\end{equation}
with $\mathbf{\mathcal{H}}$ given by
\begin{equation}
\mathcal{H}_{ij}=\langle\phi_i|S^\dagger H S |\phi_j\rangle = \langle\phi_i| H S^\dagger S |\phi_j\rangle = \sqrt{3!}\langle\phi_i| H S |\phi_j\rangle,
\end{equation}
and $\mathbf{\mathcal{N}}$ defined to be
\begin{equation}
\mathcal{N}_{ij}=\langle\phi_i|S^\dagger S |\phi_j\rangle = \sqrt{3!}\langle\phi_i| S |\phi_j\rangle.
\end{equation}
This yields the coefficients $c_i$ that minimize the energy.

The resulting binding energy, using a basis of 29 wavefunctions, is
\begin{equation}
E_b\approx -0.475\,m_{b'}\left(\alpha_s^{\prime 2} -\frac{\alpha^\prime\alpha_s^\prime}{3}+\frac{\alpha^{\prime 2}}{36}\right)~.
\end{equation}
For comparison, with $\alpha^\prime=1/137$ and $\alpha_s^\prime=0.15$, we find a binding energy of $E\approx-0.0105m_{b^\prime}$, which is a 20\% deeper than the naive estimate. 
The mass of the $\Delta'$ baryon is
\begin{equation}
m_{\Delta'}=3m_{b'}+E_b\approx m_{b'}\left[3-0.475 \left(\alpha_s^{\prime 2} -\frac{\alpha^\prime\alpha_s^\prime}{3}+\frac{\alpha^{\prime 2}}{36}\right) \right].
\end{equation}
Here the running couplings should be evaluated at the scale of typical momentum transfer within the baryon, which is defined by
\begin{equation}
\mu\equiv\alpha_s'(\mu)m_{b'}.\label{e.muScale}
\end{equation}

\begin{figure}[t]
 \begin{center}
\includegraphics[width=0.8\textwidth]{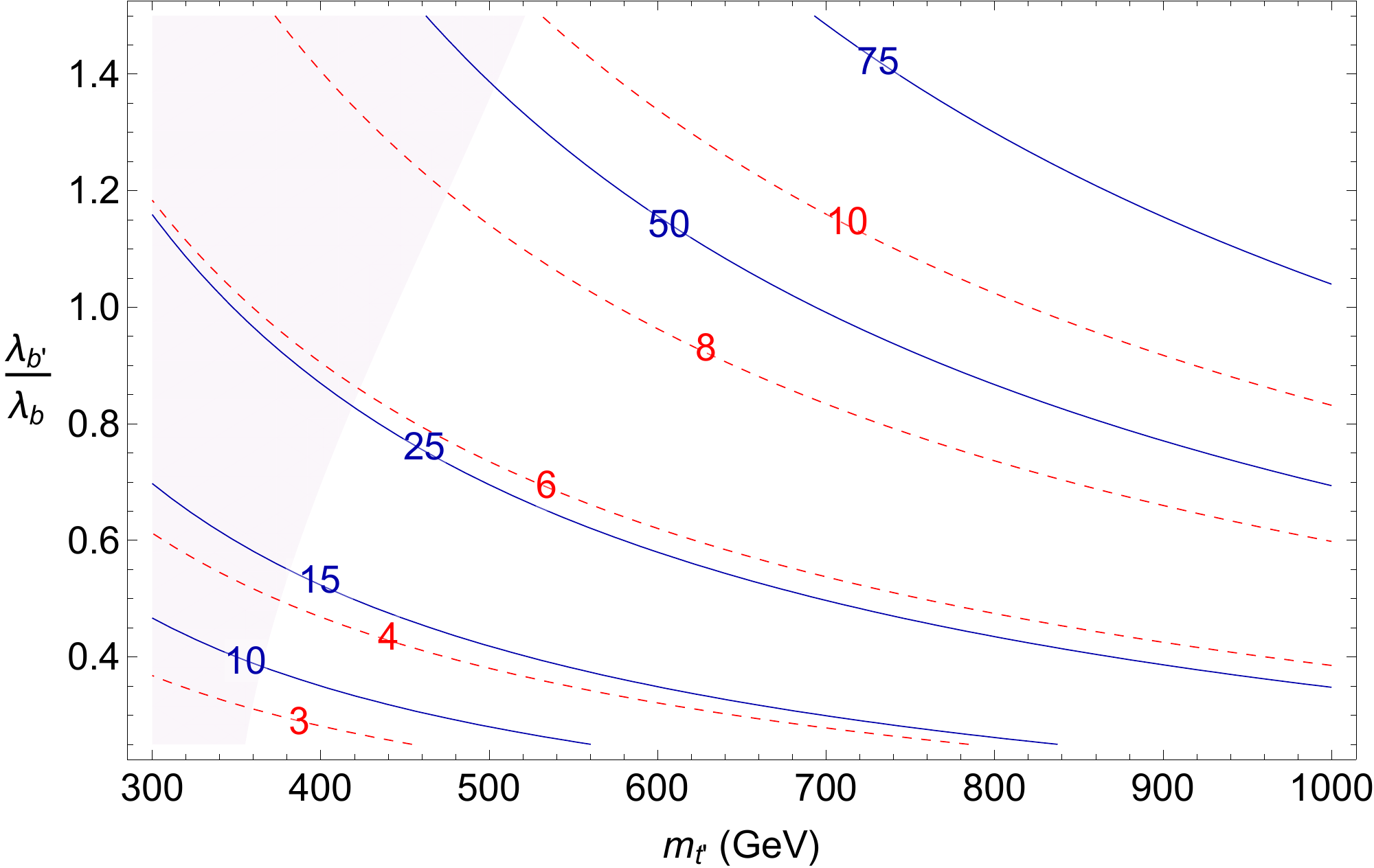} 
 \end{center}
 \caption{Contours of the $\Delta'$ baryon mass in GeV (solid blue) and the ratio $\mu/\Lambda_{\text{QCD}'}$ (dashed red) as a function of $m_{t'}$ and the ratio of the twin bottom Yukawa to the SM value. The definition of $\mu$ is given in Eq.~\eqref{e.muScale}. The shaded purple area is in tension with Higgs coupling measurements. }
\label{f.baryMass}
\end{figure}

This calculation of the binding energy assumed a Coulombic potential, but corrections from the confinement potential are of order $\Lambda_{\text{QCD}'}^2 /\mu^2$. The full potential may be parameterized as
\beq
V_\text{Total}=V_\text{Cou}+V_\text{Conf}=\frac23\frac{\alpha_s'}{r}+c\Lambda_{\text{QCD}'}^2r,
\eeq 
where $c$ is an order one number obtained from nonperturbative physics. For the Coulomb term to dominate we need 
\beq
\Lambda_{\text{QCD}'}^2\ll \frac{2}{3} \alpha_s'(\mu) \mu^2 ~.
\label{criteria}
\eeq
To remain in the regime of perturbativity we also require $\alpha_s'(\mu)<0.5$. In Fig.~\ref{f.baryMass} we plot contours of the $\Delta'$ mass as a function of the twin top-quark mass $m_{t'}$ and the ratio of the twin bottom-quark Yukawa to the SM value.  In the plot we choose the UV cutoff $\Lambda_\text{UV}$ to be 5 TeV, and take
\beq
 \frac{g'_3(\Lambda_\text{UV})-g_3(\Lambda_\text{UV})}{g_3(\Lambda_\text{UV})}\equiv\delta g_3'=-0.15\,,\label{e.deltag3}
\eeq
 to ensure we remain within the perturbative regime over the entire parameter space. As the figure makes clear, the baryon masses that best match a naive realization of ADM, that is with $m_\text{DM}\sim 5 m_N$, occur for smaller $\lambda_{b'}$ and smaller $m_{t'}$. We also note that if the bottom Yukawa is reduced too much, we violate (\ref{criteria}). We avoid these situations by taking $\lambda_{b'}/\lambda_{b}>0.15$ . The dashed red lines of the figure indicate contours of $\mu/\Lambda_{\text{QCD}'}$, which control the size of corrections to our numerical calculation. 

The figure also shades in the area in tension with Higgs coupling measurements. These regions of the parameter space reduce the rates of Higgs production and decay into visible states below 80\% of the SM prediction, which is the current bound: see Refs.~\cite{Burdman:2014zta,Craig:2015pha,Kilic:2018sew}. Therefore we find that the allowed range of twin baryon masses is about 10\textendash100 GeV. Comparing this to the requirements of ADM in Eq.~\eqref{e.adm} we find that the twin baryon asymmetry $\eta_\text{DM}$ needs to be between $\eta_\text{B}/2$ and $\eta_\text{B}/20$.


\section{Twin Baryon Dark Matter\label{s.Bconst}}
The spectrum of the fraternal twin sector is determined largely by the masses of the twin $b'$ quark, $\tau'$, and $\tau'$ neutrino. While the framework accommodates many choices, we focus on twin masses close to their SM partners. That is, we assume both the $b'$ and $\tau'$ to have masses of at least a few GeV, while the $\nu_\tau'$ is much lighter. In this case the twin $W^{\prime\pm} $ decays quickly to the lighter leptons, which are stable. 

The hadron spectrum is composed of $\overline{b'}b'$ mesons, $b'b'b'$ baryons, and glueballs. These last have masses set by the twin confinement scale, varying from $7\Lambda_{\text{QCD}'}$ to $18\Lambda_{\text{QCD}'}$. The glueballs and mesons decay quite rapidly through twin weak interactions into twin neutrinos and lighter glueball or meson states. A few of these states also mix with the Higgs, allowing for decays into SM final states. The lightest baryon state, the spin 3/2 $\Delta'$ with mass $m_{\Delta'}\sim 3m_{b'}$, is stabilized by its nonzero twin baryon number. 

In the fraternal twin Higgs setup the twin hypercharge may or may not be gauged. From pure $\Delta N_\text{eff}$ considerations it is better to have no massless degrees of freedom in the dark sector. In such a scenario the leptons are largely unimportant as far as DM signals go, and their dynamics and signals have been discussed elsewhere~\cite{Garcia:2015loa,Craig:2015xla,Garcia:2015toa}. For the remainder of this section we turn to the constraints on the twin baryons as DM.

The self-scattering cross section of $\Delta'$ baryons can be estimated from their mutual long range van der Waals potential:
\beq
V_{\rm vdW}(r)\approx c_{\rm vdW} \frac{\alpha_s'}{\mu^5} \frac{1}{r^6},
\eeq
for few $\cdot\, \Lambda_{\text{QCD}'}< r^{-1} < \mu$, where $c_{\rm vdW}$ is an order one number and $\mu$ is the characteristic momentum scale of the quarks given in Eq.~\eqref{e.muScale}. For $r> \Lambda_{\text{QCD}'}^{-1}$ there is no potential between two baryons since there are no light twin mesons to exchange.
The cross-section is then given by \cite{Massey}
\beq
\sigma \approx\frac{9\pi}{4} \left(\frac{3\pi\alpha_s' c_{\rm vdW}m_{\Delta'}}{8\mu^5 k}\right)^{2/5}~,
\eeq
where $\mu > k >\Lambda_{\text{QCD}'}$ is the momentum transfer.
The scattering cross section for DM is bounded \cite{Harvey:2015hha} to be less than 
\begin{equation}
\frac{\sigma}{m_\text{DM}}\lesssim 0.47\frac{\text{cm}^2}{\text{g}}=\left(13\,\text{GeV}^{-1}\right)^3~,
\label{sigmabound}
\end{equation}
for the DM to be consistent with observations, primarily that of the Bullet Cluster \cite{Clowe:2006eq}. To obtain a conservative bound we take the smallest value of $k=\Lambda_{\text{QCD}'}$  and the largest coupling value, $\alpha'_s=0.5$ in the numerator. So the cross section bound is satisfied for sufficiently heavy $b'$s:
\beq
 \left(\frac{\mu^5\Lambda_{\text{QCD}'}m_{b'}^{3/2}}{c_{\rm vdW}} \right)^{2/15} \gtrsim 0.11\, {\rm GeV}~.
 \label{inequ}
\eeq
Going to the edge of the interesting parameter space, $m_{t'}=300$ GeV, $\lambda_{b'}/\lambda_b=0.15$, and taking a large van der Waals coefficient of $c_{\rm vdW}=100$ gives\footnote{In light atoms the analogous values of $c_{\rm vdW}$ range from 0.25 to 150.} a value of $0.3$ $\text{GeV}$ for the left-hand side of (\ref{inequ}), well above the bound. For larger $m_{t'}$ and $\lambda_{b'}$ the bound is even more easily satisfied.


Direct detection of twin baryon DM is dominated by Higgs exchange. This contrasts with~\cite{Garcia:2015toa} which also needed to consider  $b'$ meson exchange, which can dominate in the strong coupling regime. With the $b'$ masses above the confinement scale, however, the effective coupling between the baryon $\Delta'$ and the mesons is small and the mesons are heavy. Then, the leading effect is simply from $t$-channel Higgs exchange between a target nucleon and twin baryon.

We parametrize the coupling of the Higgs $h$ to the nucleon $N$ as
\begin{equation}
\frac{m_N}{v}f_Nh\overline{N}N,
\end{equation}
where $f_N$ is calculated using lattice measurements of matrix elements of quark mass terms in the nucleon and their Yukawa couplings. In particular we use the definition
\beq
\langle N|m_q q\overline{q}|N\rangle\equiv m_N f^N_{T_q}.
\eeq
For a coupling $\lambda_q$ between the Higgs and the quarks, $\lambda_q h q\overline{q}$, the general Higgs coupling $F_N h N\overline{N}$ to the nucleon is
\beq
F_N=\sum_q \frac{\lambda_q}{m_q}m_N f^N_{T_q}=\frac{m_N}{v}\sum_q \frac{m_q}{m_q}f^N_{T_q}\equiv \frac{m_N}{v}f_N.
\eeq
Using the values in~\cite{Ellis:2018dmb} we find $f_N\approx 0.3$. However, extraction of these parameters from recent experiments may point to slightly smaller values~\cite{Alarcon:2011zs,Alarcon:2012nr}; see also~\cite{Giedt:2009mr,Low:2011kp,Crivellin:2013ipa}. In this analysis we have neglected the $\sim v^2/f^2$ correction to the Higgs-quark couplings, which would slightly decrease the final cross section. 

The coupling between the Higgs and the $\Delta'$ baryon is similarly defined,
\begin{equation}
\frac{m_{\Delta'}}{f}\frac{v}{f}f_{\Delta'} h\overline{\Delta'}\Delta',
\end{equation}
to leading order in $v/f$. This leading factor of $v/f$ comes from the Higgs coupling to the $b'$, which is $v/f$ suppressed relative to the SM coupling. Therefore, to leading order in $v/f$ and the velocity of the DM the baryon-nucleon cross section is
\begin{equation}
\sigma_{N\Delta}\approx\frac{\mu_{N\Delta}^2}{\pi m_h^4}\frac{\left(f_N m_N\right)^2}{v^4}\frac{\left(f_{\Delta'} m_{\Delta'}\right)^2}{(f/v)^4},
\end{equation}
where
\beq
 \mu_{N\Delta}=\frac{m_N m_{\Delta'}}{m_N+m_{\Delta'}}\,,
\eeq
 is the reduced mass of the nucleon-baryon system. Unlike the $m_{b'}\ll\Lambda_{\text{QCD}'}$ limit, we saw in Sec.~\ref{s.barymass} that the gluon contribution to $m_{\Delta'}$ is small. Thus, the coupling of the Higgs to the twin baryon is simply mediated by the coupling to the constituent quarks. Consequently, to leading order in $\alpha_s'$ we find $f_{\Delta'}=1$.

\begin{figure}[t]
 \begin{center}
\includegraphics[width=0.8\textwidth]{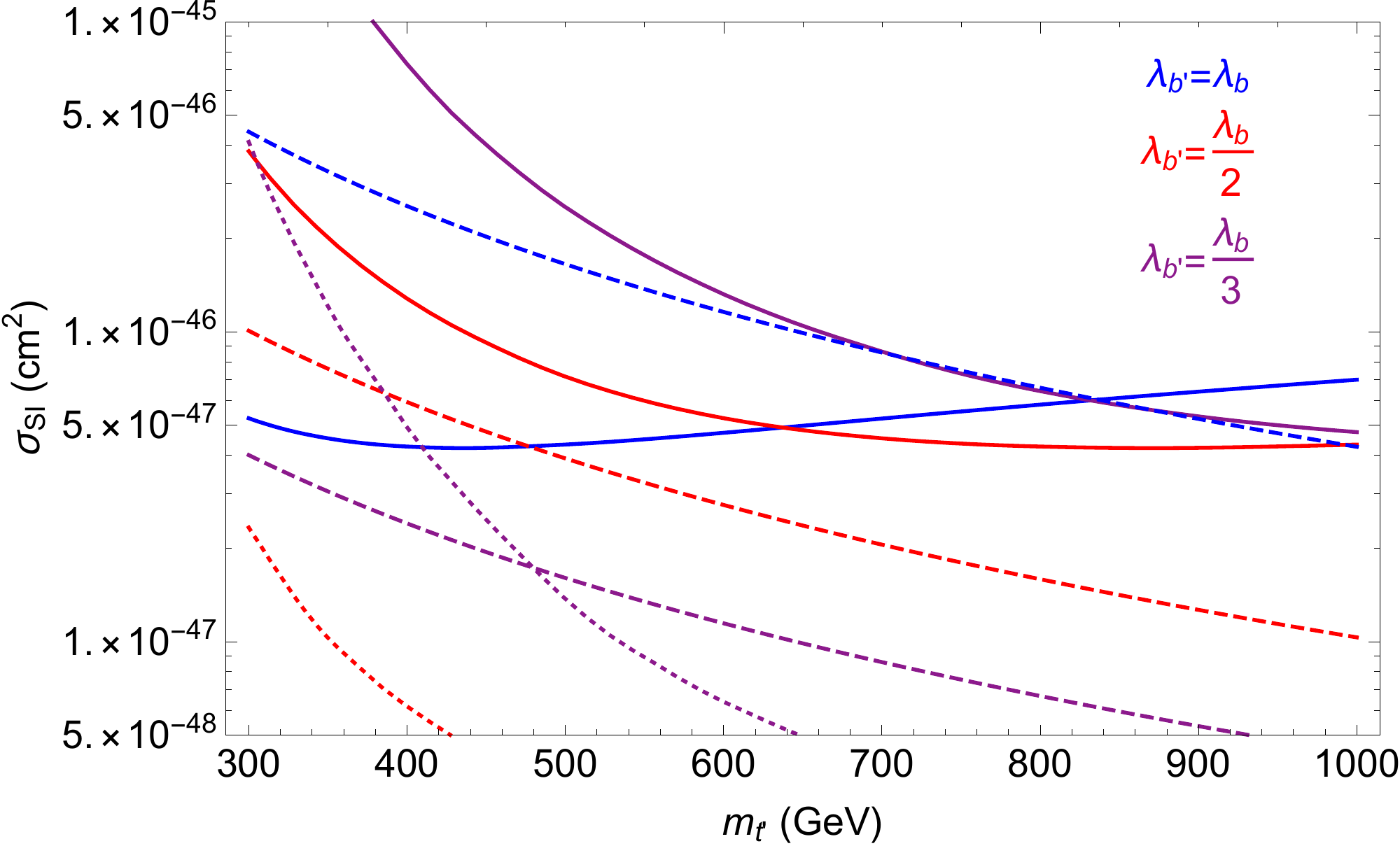} 
 \end{center}
 \caption{Plot of the direct detection cross section (dashed) and experimental bounds (solid) as a function of the twin top-quark mass $m_{t'}$. Curves are shown for twin bottom Yukawa couplings equal to 1 (blue), 1/2 (red), and 1/3 (purple) times the SM value. Dotted lines correspond to the projected sensitivity of the LZ detector.}
\label{f.DDbounds}
\end{figure}

This allows us to plot the direct detection sensitivity as a function of $m_{b'}$. In Fig.~\ref{f.DDbounds} we compare these sensitivities to the latest results from Xenon 1 Ton~\cite{Aprile:2018dbl} for several values of the twin bottom-quark Yukawa coupling $\lambda_{b'}$. The experimental bounds are a function of the $\Delta'$ mass, which is itself a function of $m_{t'}$ and $\lambda_{b'}$. For each value of the Yukawa coupling we plot both the experimental bounds (solid line) and the projected signal (dashed line) as a function of the twin top-quark mass. We see that when the SM and twin Yukawas are equal only larger values of $m_{t'}$ are consistent with experiment. However, as the twin bottom Yukawa is decreased, the projected signal falls below the current bounds. 

Recall from Fig.~\ref{f.baryMass} that larger values of $m_{t'}$ also lead to larger $m_{\Delta'}$. In addition, it is clear that if $\lambda_{b'}>\lambda_b$ then even larger values of $m_{t'}$ and $m_{\Delta'}$ would be required to agree with experiment. Thus, direct detection and naturalness (preferring lighter $m_{t'}$) push us toward twin bottom Yukawas that are smaller than the SM value. This, in turn, reduces $m_{\Delta'}$, pushing it toward the naive ADM expectation of $\sim$5 GeV. 

In short, the $\Delta'$ baryon is a successful ADM candidate, whose mass and scattering cross section can be determined with some precision. What is more, while the direct detection cross sections are smaller than the current limits, most are accessible to the proposed LZ~\cite{Akerib:2018lyp} experiment, whose projected sensitivities are given by the dotted lines in Fig.~\ref{f.DDbounds}. Only the lightest mass states can escape detection there, but the high luminosity LHC run is expected to, at least indirectly, probe these states up to $m_{t'}=500$ GeV through the corresponding modifications to Higgs couplings~\cite{Kilic:2018sew}. 


\section{Twin Atom Dark Matter\label{s.DarkAtoms}}
If the twin $U(1)_{Y'}$ is gauged, then twin atoms composed of $\tau'$ particles bound to $\Delta'$s typically result. This is ensured by twin charge neutrality: the asymmetric production of $\Delta'$ requires a compensating asymmetry in $\tau'$. The mass of the dark atom $m_D$ is simply
\beq
m_D=m_{\Delta'}+m_{\tau'}-B_D,
\eeq
where $B_D$ is the binding energy, which is, to leading order,
\beq
B_D= \frac{\alpha^{\prime 2}\mu_D}{2},
\eeq
where $\mu_D$ is the reduced mass of the $\Delta'$-$\tau'$ system and $\alpha'$ is the twin fine structure constant. Making use of the definition $R=m_{\Delta'}/m_{\tau'}$ we can then express the reduced mass (and therefore the binding energy) as
\beq
\mu_D\approx\frac{2m_D R}{2(1+R)^2-\alpha'^2R}\,.
\eeq
These relations are used repeatedly below.

If the recombination of these particles into twin atoms is not sufficiently efficient then the DM remains primarily a plasma, which can develop instabilities that affect galaxy collisions, like the Bullet Cluster \cite{Clowe:2006eq}. This translates into a bound on the twin fine structure constant $\alpha'$ as a function of $m_D$~\cite{CyrRacine:2012fz}:
\begin{equation}
\frac{\alpha^{\prime 4}}{\xi}\left(\frac{\Omega_D h^2}{0.11} \right)\left( \frac{\text{GeV}}{m_D}\right)^2\left[\frac{(1+R)^2}{R}-\frac12\alpha^{\prime 2} \right]^2\gtrsim 7.5\times 10^{-11},\label{e.plasmaBound}
\end{equation}
where $\Omega_D h^2$ is the relic density of dark matter and $\xi$ is the ratio of the present day temperature of the dark radiation to the CMB temperature
\begin{equation}
\xi=\left. \left(\frac{T_D}{T_\text{CMB}} \right)\right|_{z=0}.
\end{equation}

We determine $\xi$ in steps. The Higgs portal keeps the two sectors in thermal equilibrium down to a decoupling temperature $T_\text{dec}$ of a few GeV~\cite{Chacko:2016hvu}. After decoupling the two sectors evolve independently, each conserving entropy. This allows us to relate the ratio of temperatures today $\xi$ to the ratio at decoupling, which is $\xi_\text{dec}=$1:
\begin{equation}
\xi=\xi_\text{dec}\left( \frac{g_\ast^\text{today}g_{\ast D}^\text{dec}}{g_\ast^\text{dec}g_{\ast D}^\text{today}}\right)^{1/3}.
\end{equation}
 Here $g_{\ast(D)}$ is the effective number of degrees of freedom in the visible (dark) sector
 \begin{equation}
 g_\ast=\sum_\text{bosons}g_i+\frac78\sum_\text{fermions}g_i,
 \end{equation}
 where the sum is over all relativistic degrees of freedom at a given temperature. For instance, at decoupling $g^\text{dec}_\ast=75.75$ while $g_\ast^\text{today}=3.94$ is the present value. If the two sectors decouple before the twin QCD phase transition, then $g_{\ast D}^\text{dec}$ includes contributions from the twin gluons, making $\xi$ much larger than otherwise. In other words, if the sectors decouple before the twin phase transition, then the twin photon and light leptons receive all the entropy from the phase transition, and their final temperature is correspondingly higher. By simply choosing $\delta g_3'$, as defined in Eq.~\eqref{e.deltag3}, to be nonnegative the twin confinement scale is $\gtrsim5$ GeV~\cite{Craig:2015pha}, comfortably above the decoupling temperature.
 
The masses of the twin sector particles $b'$, $\tau'$, and $\nu'_\tau$ are not fixed, which affects the temperatures at which they are relativistic degrees of freedom. However, the $b'$ is typically too heavy to contribute much at $T_\text{dec}$. Then, the largest value of $\xi$ results from assuming both the $\tau'$ and the $\nu_\tau'$ contribute at decoupling, but neither do today. With these assumptions $\xi\approx 0.57$, which is well within the bounds on new relativistic degrees of freedom at BBN~\cite{CyrRacine:2012fz}, but we must be more careful about the CMB bounds.

The energy density $\rho$ at CMB times can be written as
\beq
\rho_\text{CMB}=\frac{\pi^2}{15}T_\gamma^4\left[ 1+\frac78\left( \frac{4}{11}\right)^{4/3}N_\text{eff}^\text{SM}+\frac78\left( \frac{4}{11}\right)^{4/3}\Delta N_\text{eff}\right],
\eeq
where $T_\gamma$ is the temperature of the visible photons. Assuming the twin $\tau'$ is nonrelativistic at these energies, and that the twin photon and twin neutrino $\nu_\tau'$ have the same temperature we obtain
\beq
\Delta N_\text{eff}=\xi^4_\text{CMB}\left[\frac87\left( \frac{11}{4}\right)^{4/3}+N_{\nu'} \right],
\eeq
where $N_{\nu'}$ counts the number of active neutrino species at CMB energies. For $N_{\nu'}=0$ we have $\xi\approx 0.57$ and $N_\text{eff}\approx0.48$, in tension with the 2$\sigma$ bound of $\Delta N_\text{eff}<0.3$. However, if the twin neutrino is still active at CMB times then $\xi\approx 0.465$ and $N_\text{eff}\approx0.25$ which agrees with current measurements. Thus, for the remaining bounds we take $\xi=0.465$.
 
 \begin{figure}[t]
 \begin{center}
\includegraphics[width=0.8\textwidth]{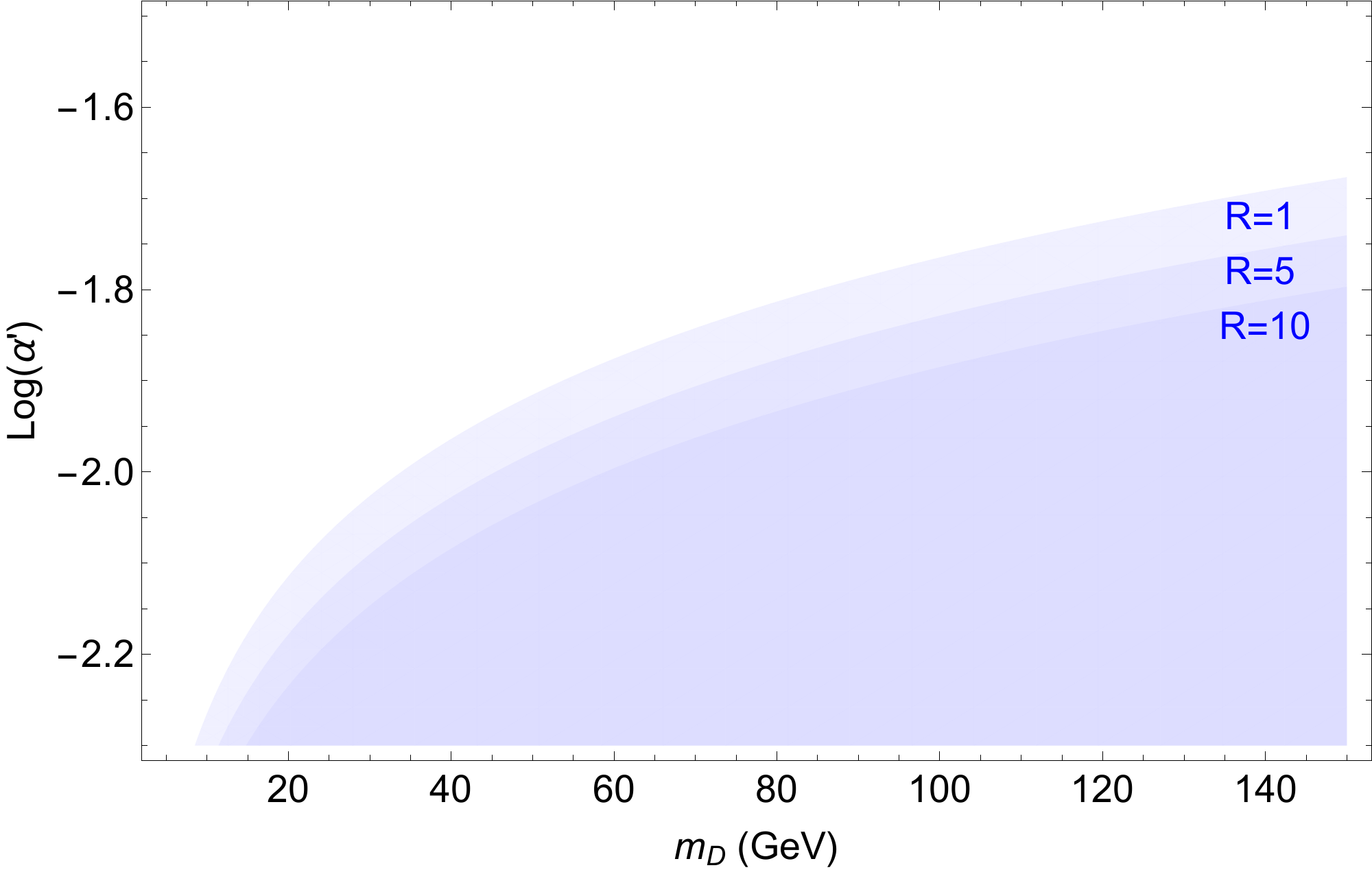} 
 \end{center}
 \caption{Exclusion regions for dark atoms from dark plasma stability as a function of the dark atom mass $m_D$ and the logarithm dark fine structure $\alpha^{\prime}$. Three benchmarks of $R=m_{\Delta'}/m_{\tau'}$ are shown. }
\label{f.darkAtom}
\end{figure}
 
The bounds on the twin dark atom scenario from Eq.~\eqref{e.plasmaBound} are shown in Fig.~\ref{f.darkAtom} for three values of $R$. The figure makes clear that twin DM with lighter masses is less constrained. Also a larger mass hierarchy between $m_{\Delta'}$ and $m_{\tau'}$ weakens the constraint. If the twin QED coupling is similar to the SM value $\alpha'\lesssim1/100$, then only lighter DM masses agree with the data. 


A second bound on twin atoms comes from their self interactions. Again, from \cite{Harvey:2015hha} one finds that their self-interaction cross section must satisfy (\ref{sigmabound}). The self-scattering cross section of twin atoms can be estimated from their mutual long range van der Waals potential:
\beq
V_{\rm vdW}(r)\approx c_{\rm vdW,a} \alpha' \frac{a^5}{r^6},
\label{vdWatom}
\eeq
where $c_{\rm vdW,a}$ is again an order one number and $a$ is the twin Bohr radius:
\beq
a=\frac{1}{\alpha' \mu_D}~.
\eeq
The cross-section is then given by \cite{Massey}
\beq
\sigma \approx\frac{9\pi}{4} \left(\frac{3\pi\alpha' c_{\rm vdW,a} m_D a^5}{8 k}\right)^{2/5}~,\label{e.atomXsec}
\eeq
where $k$  is the momentum transfer.

\begin{figure}[t]
 \begin{center}
\includegraphics[width=0.8\textwidth]{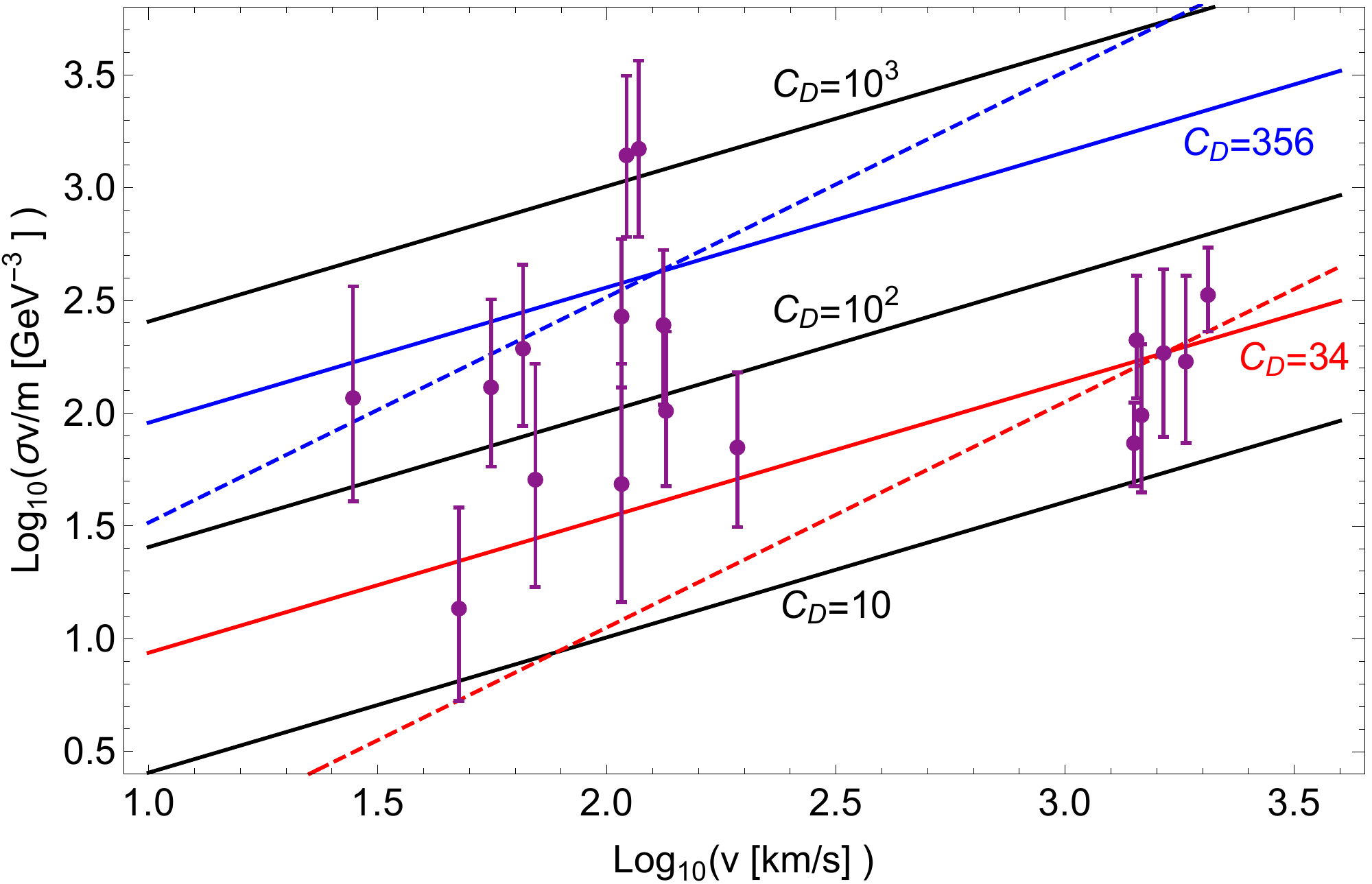} 
 \end{center}
 \caption{Plot of the twin atom self interaction cross section divided by DM mass and multiplied by velocity as a function of velocity for several values of $C_D$ in units of $\text{GeV}^{-3}$. Data from dwarf galaxies and low surface brightness spiral galaxies are clustered at lower velocities with galaxy cluster measurements at higher velocities. In red (blue) we plot the best fit line appropriate to the galaxy cluster (galaxy only) data. These are compared to similar best fit lines corresponding to velocity independent cross sections. }
\label{f.darkAtomVel}
\end{figure}

It is possible that the velocity dependence of this cross section can solve a minor problem with standard DM candidates \cite{Chu:2018faw,Kaplinghat:2015aga}. In order to make this comparison we use Eq.~\eqref{e.atomXsec} to obtain
\beq
\frac{\sigma v_D}{m_D} \approx v_D^{3/5}\frac{9\pi}{16}\left( \frac{3\pi}{8}\right)^{2/5}C_D\,.
\eeq
where we have defined
\beq
C_D=\left(\frac{c_{\rm vdW,a}}{\alpha'^4} \right)^{2/5}\frac{\left[2(1+R)^2-\alpha'^2R \right]^2}{m_D^3R^2}.
\eeq
In Fig.~\ref{f.darkAtomVel} we  plot this cross section as a function of velocity for several values of $C_D$ in units of $\text{GeV}^{-3}$. Also shown are the data given in~\cite{Chu:2018faw,Kaplinghat:2015aga} obtained from dwarf and low-surface-brightness galaxies (at lower velocities) and galaxy clusters (at higher velocities). 

The figure shows separate best fit values for the low and high velocity data, which are $C_D=356$ $\text{GeV}^{-3}$ and $C_D=34$ $\text{GeV}^{-3}$ respectively. These lines are contrasted with similar best fit lines for DM with a velocity independent cross section. Taking the high velocity data as somewhat more consistent, we see that twin atom DM provides a better agreement with the low velocity data than DM with velocity independent cross sections. A similar velocity dependence would arise in any composite dark matter model \cite{Alves:2009nf,Kribs:2009fy,Frigerio:2012uc,Wise:2014jva} that has long range dipole-dipole interactions as in Eq. (\ref{vdWatom}).

\begin{table}
\begin{center}
\begin{tabular}{| c | c | c | c |}
\hline
$c_{\rm vdW,a}$ & $\alpha'$ & $m_D$ & $R_\text{Min}$ \\
\hline 
\hline 
1 & 1/18 & 5 GeV & 1.4
 \\
\hline
1 & 1/137 & 15 GeV & 1.6
 \\
\hline
10 & 1/10 & 5 GeV & 1.5
 \\
\hline
10 & 1/137 & 20 GeV & 1.3
 \\
\hline
\end{tabular}
\end{center}
\caption{\label{t.benchmarks}Benchmark points near the real, positive $R>1$ threshold with $C_D=55$ $\text{GeV}^{-3}$. $R$ can be increased by increasing either $\alpha'$ or $m_D$.}
\end{table}

We can also use the plot to determine how the physical parameters, rather than the combination in $C_D$, must be related. The best fit to all the experimental data is $C_D\sim 55$ $\text{GeV}^{-3}$. Assuming this value we relate the remaining parameters. Specifically, by specifying $c_{\rm vdW,a}$ we can explore how $\alpha'$ and $m_D$ determine $R$. We provide some benchmarks in Table~\ref{t.benchmarks}, which are taken near the boundary where $C_D=55$ $\text{GeV}^{-3}$ can be solved for $R>1$. For instance, taking the naive ADM benchmark of $m_D=5$ GeV and $c_{\rm vdW,a}=1$ we must take $\alpha'\sim1/15$ to obtain a real positive $R$, and find $R=1.4$. Increasing $\alpha'$ leads to larger $R$. On the other hand, for $\alpha'=1/137$ and $c_{\rm vdW,a}=10$ we must take $m_D=20$ GeV to find a physical value for $R$, in this case $R=1.3$, with larger $R$ resulting from larger $m_D$. However, increasing $m_D$ leads to tension with the recombination bounds shown in Fig.~\ref{f.darkAtom}. 

The above analysis applies to elastic scattering of the twin atoms. However, it has been shown~\cite{Boddy:2016bbu} that the hyper-fine splitting of the ground state can lead to inelastic scattering. This provides additional velocity dependence to the DM scattering which can explain some of the questions about large-scale structure, see~\cite{Prilepina:2016rlq} for another application of these results in the twin Higgs framework. During inelastic collisions the atom is up-scattered into this excited hyperfine state, which then decays by emitting hidden photons. This process can be important when the kinetic energy of the dark atoms is similar to the splitting between the hyperfine states. To estimate this splitting we simply adapt the standard result for Hydrogen hyperfine splitting to our case with a spin-3/2 ``proton." We find the energy splitting between the spin-2 states and spin-1 states is
\beq
\Delta E_\text{hf}=\frac83 g_\Delta\alpha'^4\frac{m_{\tau'}^2}{m_{\Delta'}}=\frac83\frac{g_\Delta \alpha'^4m_D(1+R)}{R(1+R)^2-\alpha'^2R^2/2},
\eeq
 where $g_\Delta$ is the Land\'{e} $g$-factor of the $\Delta'$ and in the perturbative limit we are considering should be close to six, the sum of the three $g_{b'}$. Note that this is a factor of four larger than the Hydrogen like case, because the twin baryon is spin-3/2. Then, by setting this equal to the kinetic energy $\frac12 m_D v^2$, we find the velocities $v_\text{In}$ for which this type of scattering is important:
 \beq
 v_\text{In}^2\sim \frac{16}{3}\frac{g_\Delta \alpha'^4(1+R)}{R(1+R)^2-\alpha'^2R^2/2}~.\label{e.elastVel}
 \eeq
 
  \begin{figure}[t]
 \begin{center}
\includegraphics[width=0.8\textwidth]{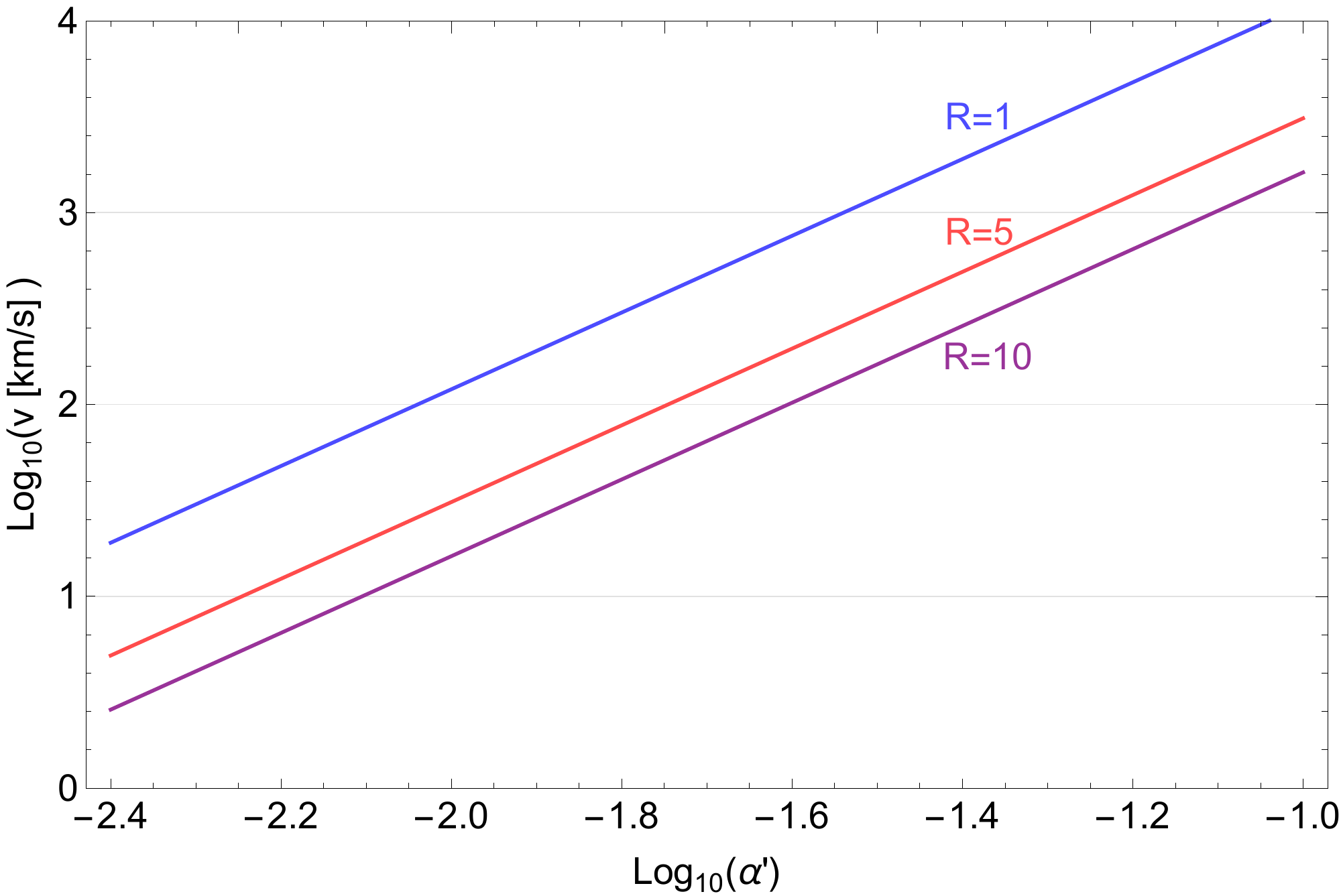} 
 \end{center}
 \caption{Plot of the velocity when inelastic collisions to the hyperfine split excited ground state are maximized as a function of the twin fine structure constant for several values of $R$. We take $g_\Delta=6$.  }
\label{f.inElastVel}
\end{figure}

In Fig.~\ref{f.inElastVel} we use Eq.~\eqref{e.elastVel} to find the velocities at which the inelastic collisions to the hyperfine excited state are most important. We see that there are parameter values where these collisions can further modify the velocity dependence of the self-scattering cross section. By comparing with Fig.~\ref{f.darkAtomVel} we note that as the ratio in particle masses $R$ becomes larger the twin $U(1)$ coupling must become larger in order for the velocities associated with galaxy clusters to be affected. The typical velocities in the individual galaxy data can be modified when the twin coupling is closer to the SM value.

Before dark recombination the baryon-photon fluid in the twin sector undergoes acoustic oscillations. These can in turn leave traces on the visible matter power spectrum, affecting large scale structure. A detailed study of such dark acoustic oscillations~\cite{Cyr-Racine:2013fsa} encapsulates these effects through the parameter $\Sigma_\text{DAO}$. In our case we estimate this quantity to be
\beq
\Sigma_\text{DAO}\approx\frac{2(1+R)^2-\alpha'^2R}{\alpha' R}10^{-9}\left( \frac{\text{GeV}}{m_D}\right)^{7/6}\,.
\eeq
Since the constraints in Fig.~\ref{f.darkAtom} require $\alpha'\gtrsim10^{-2}$, we find $\Sigma_\text{DAO}<10^{-7}$, which is far below the sensitivity of current analyses.

Finally, the direct detection of these twin atoms is qualitatively similar to the baryon only case. There are additional Higgs exchange effects from the twin $\tau'$, but these are sub-leading to the baryon-baryon interactions, unless the twin $\tau'$ Yukawa coupling is raised considerably. In effect, this is controlled by $R$, since both the baryon and the $\tau'$ get much of their mass from the Higgs. So, for $R$ somewhat larger than one the bounds in Fig.~\ref{f.DDbounds} should be approximately correct.

In short, twin atoms can make up an interesting ADM population. To have $m_D$ values closest to 5 GeV, the $\tau'$ mass should be close to $m_{\Delta'}$, so that $R\sim1$. These lightest mass atoms also require the $\alpha'$ coupling be somewhat stronger than in the visible sector. In addition, the velocity dependence of the self-interaction of these twin atoms agrees with self-interaction estimates better than DM with a velocity independent self-interaction cross section.


\section{Conclusions\label{s.conc}}
The fraternal twin Higgs scenario provides simple asymmetric DM candidates while stabilizing the Higgs mass up to TeV scales. We have demonstrated that twin baryons, whose constituent quarks have masses above the twin confining scale, successfully realize the ADM construction in a simple way. This is true both when the twin $U(1)$ is gauged and when it is not. 

The most compelling regions of parameter space have lighter quarks and baryons, which also reduces the direct detection signal. Thus, the observation that $\Omega_DM\simeq 5\,\Omega_B$ is correlated with direct detection signals below the current state of the art. However, projected next generation sensitivities cover nearly all the motivated parameter space. The remaining parameter range will be indirectly probed by Higgs coupling measurements at the LHC. Other LHC searches, including for displaced vertices associated with boson decays, provide additional experimental tests of the fraternal twin Higgs set-up. In the case of twin atoms, it would be interesting to work out how much clumping can occur as a result of inelastic scattering.

\section*{Acknowledgments}
We thank Zackaria Chacko, Hsin-Chia Cheng, and Isabel Garc\'{i}a Garc\'{i}a for helpful discussions.  This research is supported in part by DOE grant DE-SC-0009999.

\end{document}